\documentclass[twocolumn]{autart}
\usepackage[utf8]{inputenc} 
\usepackage[T1]{fontenc}    
\usepackage{url}            
\usepackage{booktabs}       
\usepackage{amsfonts}       
\usepackage{nicefrac}       

\usepackage{graphicx}
\usepackage{subfigure}
\usepackage{caption}
\usepackage{indentfirst}
\usepackage{placeins}
\usepackage{float} 
\usepackage{svg}
\usepackage{epstopdf} 
\usepackage{algorithm}
\usepackage{algpseudocode}

\usepackage{amsmath}
\usepackage{amssymb}
\allowdisplaybreaks[2]
\usepackage{graphicx}          
\usepackage{amsmath}
\usepackage{amssymb}
\usepackage{float} 

\newtheorem{lemma}{Lemma}
\newif\ifqedhere
\qedherefalse

\newenvironment{proof}
{{\noindent\it Proof.}\ }
{%
  \ifqedhere
    \global\qedherefalse
  \else
    \hfill $\square$
  \fi
  \par
}

\newcommand{\myqedhere}{%
  \tag*{\(\square\)}%
  \global\qedheretrue
}
\usepackage{indentfirst}
\usepackage{xcolor}
\usepackage{algorithm}  
\usepackage{algpseudocode}  
\usepackage{enumerate}
\usepackage{tabularx}
\newtheorem{remark}{Remark}
\newtheorem{theorem}{Theorem}
\newtheorem{assumption}{Assumption}

\usepackage{footnote} 
\usepackage[colorlinks=true,allcolors=blue]{hyperref}

\begin{document}
\setlength{\parindent}{1em}
\begin{frontmatter}

\title{Learning the NMPC with Stability Guarantees\thanksref{footnoteinfo}} 
\thanks[footnoteinfo]{This paper was not presented at any IFAC
conference. 
$^{*\,}$Corresponding authors: Qifang Liu and Hong Chen.}

\author[NKL,Comm]{Shuyou Yu}\ead{shuyou@jlu.edu.cn},  
\author[Comm]{Yan Sun}\ead{sunyan25@mails.jlu.edu.cn},               
\author[Comm]{Jun Ai}\ead{aijun24@mails.jlu.edu.cn},               
\author[NKL,Comm]{Qifang Liu}\ead{liuqf@jlu.edu.cn}
\author{\hspace{-0.1em}$^{,*\,}$},
\author[Comm,TJ]{Hong Chen}\ead{chenhong2019@tongji.edu.cn}  
\author{\hspace{-0.1em}$^{,*\,}$}
\address[NKL]{National Key Laboratory of Automotive Chassis Integration and Bionics, Jilin University, China}
\address[Comm]{Department of Control Science and Engineering, Jilin University, China}  
\address[TJ]{The College of Electronics and Information Engineering, Tongji University, China}

\begin{keyword}                           
Model predictive control; Neural networks; Nonlinear systems; Learning-based control;  Stability             
\end{keyword}                             

\begin{abstract}                          
Although nonlinear model predictive control (NMPC) is widely used, its online optimization is computationally expensive. Neural network approximations of NMPC can effectively reduce the online computational burden but generally lack stability guarantees. This paper proposes an approximate NMPC framework with guarantees of constraint satisfaction and closed-loop stability. The approximation error is addressed through constraint tightening, and the resulting tightened NMPC is used to generate the training dataset of the approximate NMPC. The maximum allowable approximation error is prescribed a priori for the design of tightened NMPC and satisfied in a probabilistic sense through an iterative trial-and-error procedure alternating between network training and a posteriori verification.
Finally, the effectiveness of the proposed framework is demonstrated through a numerical example of a constrained nonlinear system.
\end{abstract}

\end{frontmatter}
\endNoHyper
\section{Introduction}
\label{section:intro}
\begin{figure*}[t]
\centering
\includegraphics[width=0.99\textwidth]{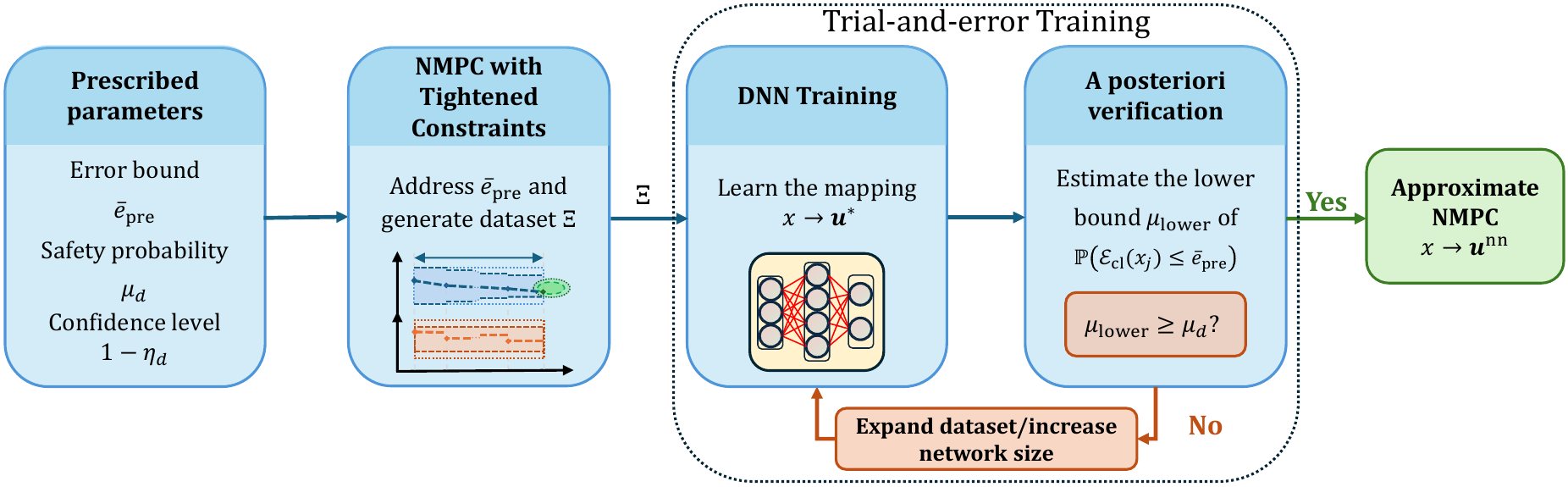}
\caption{The proposed approximate NMPC framework.}
\label{fig:Framework}

\end{figure*}
Nonlinear model predictive control (NMPC) has gained widespread application in constrained nonlinear systems due to its ability to handle constraints online, e.g. industrial processes \cite{QIN2003733}, robotics \cite{Katayama04032023} and vehicles \cite{vehiclerewiewofmpc}.

Typically, NMPC can be expressed as the following steps: (i) solving a constrained nonlinear programming (NLP) problem online, which is formulated based on the system model and initialized with the current state; (ii) applying the first control action to the plant; and (iii) repeating the entire process at the next sampling instant.

However, since constrained NLP is generally computationally expensive, it poses challenges for the online application of NMPC, particularly for tasks requiring fast and stable controller responses.
Methods for accelerating NMPC computations can be broadly classified into three categories \cite{Gonzalez20241130}: efficient solvers, explicit MPC, and neural network approximations. 

Efficient solvers, such as IPOPT \cite{WACHTER2006} and OSQP \cite{osqp}, employ advanced optimization algorithms to achieve efficient solutions to the problem. However, the optimization problem remains computationally expensive, and undesirably long computation times may still arise as the problem complexity increases.

Explicit MPC solves the optimization problem offline to derive control laws parameterized by the initial state, reducing the online computational burden to a table lookup \cite{BEMPORAD20023,JOHANSEN2004293}. However, the complexity of the explicit solution can grow rapidly with the problem dimension, and it is difficult to obtain an accurate explicit solution due to the complexity of the solution mapping of the constrained NLP.

Neural network approximation, i.e., replacing the online optimization of NMPC with a deep neural network (DNN), enables fast online computation and has attracted increasing attention \cite{VAUPEL2020261,8970557,9447448}. However, the stability of the approximate NMPC remains a key challenge due to the black-box nature of DNN.

Ensuring safety of approximate NMPC requires accounting for approximation errors introduced by DNN. Tube-based approaches address these errors by employing an ancillary robust feedback law but may involve complicated tube construction and additional design assumptions \cite{Hertneck2018}. Other approaches employ online safety mechanisms, such as verification with fallback control \cite{11159581} or projection onto robustly admissible inputs \cite{BONZANINI20205279}, potentially increasing computational complexity or affecting control performance.

An overview of the proposed method is shown in Figure \ref{fig:Framework}. The main contribution is to propose a DNN-based approximate NMPC with stability guarantees. Compared with the existing approximate NMPC framework \cite{Gonzalez20241130}, the proposed framework introduces two main improvements: (i) the combination of constraint tightening and a trial-and-error training strategy enables an a priori design of the approximation accuracy and an a posteriori verification of constraint satisfaction and stability requirements; and (ii) the proposed framework establishes explicit theoretical guarantees, including constraint satisfaction, ultimate boundedness, and asymptotic stability of the closed-loop system. Our work is inspired by the concept of inherent robustness in NMPC \cite{YU20142269}, robust NMPC \cite{Limon}, dual-mode NMPC \cite{dualmodeNMPC} and motivated by the earlier work of learning-based NMPC \cite{Hertneck2018}.

This paper is organized as follows. The problem is formulated in Section \ref{section:problem}. Section \ref{section:learning} presents the approximate NMPC framework that guarantees constraint satisfaction and stability, together with a verification scheme for the approximation error. A nonlinear constrained control benchmark example is shown in Section \ref{section:examp}. Section \ref{section:con} concludes the paper.

\subsection{Notations and Basic Definitions}
For a vector $v \in \mathbb{R}^n$, let $\|v\|$ denote the 2-norm and define the weighted norm $\|v\|^2_Q := v^\top Qv$ for $Q \in \mathbb{R}^{n\times n}$ and $Q \succeq 0$. Define the set $\mathcal{B}(\delta):=\left\{x \in \mathbb{R}^{n} \mid\|x\| \leq \delta\right\}$. Let $\mathbb{N}$ denote the set of natural numbers. For $a,b\in\mathbb{N}$ and $a\leq b$, define the integer intervals $\mathbb{I}_{a:b}:=\{a,...,b\}$. The composition of two functions $f$ and $g$ is denoted by $f \circ g(\cdot) = f (g(\cdot))$. The operation $\oplus$ is the Minkowski sum of sets $\mathcal{A} \subset \mathbb{R}^n$ and $\mathcal{B} \subset \mathbb{R}^n$, where $\mathcal{A}\oplus \mathcal{B}:=\{a+b\in \mathbb{R}^n|a \in \mathcal{A},b\in \mathcal{B}\}$. 

\section{Problem Statement}
\label{section:problem}
\subsection{Nonlinear Model Predictive Control}
\label{NMPC}
Consider a discrete nonlinear system
\begin{equation}\label{sys}
x_{k+1}=f(x_k,u_k),\quad k\geq0,
\end{equation}
where $x_k\in \mathbb{R}^{n_x}$, $u_k\in \mathbb{R}^{n_u}$ and $f: \mathbb{R}^{n_x}\times \mathbb{R}^{n_u}\rightarrow \mathbb{R}^{n_x}$ denotes the system state, control input and system dynamics, respectively.

The NMPC optimization problem \cite{MAYNE2000789,MPCbook} is formulated as follows:

\begin{prob} \label{prob1}
    \begin{align*}
    &\mathop{\mathrm{min}}\limits_{\boldsymbol {u}_k} \ J(x_k,\boldsymbol {u}_k )\\
\mathrm{s.t.}\quad
    & \overline{x}_{0|k}=x_k,\quad \overline{x}_{i+1|k}=f(\overline{x}_{i|k},u_{i|k}),\\
    & u_{i|k} \in \mathcal{U},\quad \overline{x}_{i|k} \in \mathcal{X},\quad i\in \mathbb{I}_{0:N_p-1},\\
    & \overline{x}_{N_p|k} \in \mathcal{X}_f ,
\end{align*}
\end{prob}
where $N_p$ denotes the prediction horizon and the cost function is defined as follows:
\begin{align*}
J(x_k,\boldsymbol {u}_k)&:=\sum_{i=0}^{N_p-1} l(\overline{x}_{i|k},u_{i|k})+V_f(\overline{x}_{N_p|k})\\
&= \sum_{i=0}^{N_p-1}  \left(\|\overline{x}_{i|k}\|_{Q}^{2}+\|u_{i|k}\|_{R}^{2}\right)+\| \overline{x}_{N_p|k} \|_{P}^{2},
\end{align*}
\indent The terms $\boldsymbol{\overline{x}}_k:=\{\overline{x}_{i|k}\}^{N_p}_{i=0}$ and $\boldsymbol{u}_k:=\{u_{i|k}\}_{i=0}^{N_p-1}$ represent the predicted state trajectory and the control sequence from $x_k$, respectively. Define $\boldsymbol{u}^*_k:=\{u_{i|k}^*\}_{i=0}^{N_p-1}$ as the optimal control sequence of optimization problems.

The control constraint set $\mathcal{U} \subset \mathbb{R}^{n_u}$ and the state constraint set $\mathcal{X}\subset \mathbb{R}^{n_x}$ are compact sets, with $0 \in \operatorname{int} \mathcal{U}$ and $0 \in \operatorname{int} \mathcal{X}$. The matrices $Q\in\mathbb{R}^{n_x\times n_x}$ and $R\in\mathbb{R}^{n_u\times n_u}$ are positive definite state and control weighting matrices, respectively, and $P\in \mathbb{R}^{n_x\times n_x}$ is a positive definite terminal penalty matrix. The terminal set $ \mathcal{X}_{f}:= \{x \in \mathbb{R}^{n_x} | x^\top  Px \leq \alpha\}\subseteq \mathcal{X}$ is a sublevel set of the terminal penalty function $V_f (\cdot)$.

Only the first control action $u^*_{0|k}$ is applied to the system (\ref{sys}), and the procedure is repeated at the next time instant using the new state measurement $x_{k+1}$.

In this paper, the following assumptions are made.

\begin{assumption}\label{assu:1}
The system state can be measured instantaneously and precisely.
\end{assumption}

\begin{assumption}\label{assu:2}
The system dynamic $f$ is continuously differentiable with respect to $x,u$ on $\mathcal{X}\times\mathcal{U}$ and $f(0,0)=0$. Thus, $0 \in \mathbb{R}^{n_x}$ is an equilibrium of the system.
\end{assumption}
Let $x^+=Ax+Bu$ denote the linearization of system (\ref{sys}), where $A=f'_x(0,0)$ and $B=f'_u(0,0)$. 

\begin{assumption}\label{assu:3}
The pair of $(A,B)$ is stabilizable.
\end{assumption}

By Assumption~\ref{assu:3}, there exists a feedback control law $u=Kx$ with $K\in \mathbb{R}^{n_u\times n_x}$ such that the controlled system $x_{k+1} = A_Kx_k$ is asymptotically stable, where $A_K:=(A+BK)$.
\begin{lemma}[Terminal Ingredient \cite{MPCbook}]
\label{terminalcontrollaw}
Suppose that Assumptions \ref{assu:1}-\ref{assu:3} are satisfied. Let $Q_K:=Q+K^\top RK$. Then

1) For any $\tau>1$, the Lyapunov equation
\begin{equation}\label{equ::terminalcontrollaw}
    A^\top _KPA_K+\tau Q_K=P,
\end{equation}
admits a unique positive definite solution $P$.

2) There exists $\alpha>0$ such that the terminal set \(\mathcal{X}_f\) satisfies $Kx\in\mathcal{U}$ and
\begin{align*}
V_f(f(x,Kx))-V_f(x)&\leq-x^\top Q_Kx.
\end{align*}
 for all $x\in\mathcal{X}_f$. Then, $\mathcal{X}_f$ is positively invariant for the nonlinear system (\ref{sys}) with $u=Kx$.
\end{lemma}
\subsection{Deep Neural Network}
\label{section:DNN}
Consider the mapping defined by NMPC policy: $x\in\mathbb{R}^{n_x}\rightarrow \boldsymbol {u}^*\in\mathbb{R}^{N_p \times n_u}$. The goal of the approximate NMPC is to learn the mapping through the DNN\footnote{
Since DNNs are vector-to-vector mappings \cite{DeepLearning}, the control sequence $\boldsymbol{u}^*$ is flattened into $\mathbb{R}^{N_p\cdot n_u}$ as the training target and reshaped back during the application. This procedure is omitted for simplicity.}. 
Let $M$ and $N_j$ denote the depth of the DNN and the width of layer $j\in\mathbb{I}_{1:M}$, respectively. Each layer is composed of affine-linear functions with activation functions, which can be expressed as 
\[
h_j(y_{j-1};W_j,c_j) = g ( W_j ^ { T } y_{j-1} + c_j ),
\]
where $y_{j-1}\in \mathbb{R}^{N_{j-1}}$ is the input of layer $j$. The term $g(\cdot)$ is the nonlinear activation function (e.g. ReLU \cite{relu}), while $W_j\in\mathbb{R}^{N_{j-1}\times{N_{j}}}$ and $c_j\in\mathbb{R}^{N_{j}}$ are weights and biases of the affine-linear functions.

Then the DNN can be expressed as $h(x)=h_{M}\circ h_{M-1} \circ \cdots \circ h_1(x)=\boldsymbol{u}^{\mathrm{nn}}$, where $\boldsymbol{u}^{\mathrm{nn}}$ denotes the approximate control sequence. The weights $W$ and biases $c$ of all layers are trained by minimizing the empirical loss over the training dataset $\Xi$ to accurately approximate the mapping $x\rightarrow \boldsymbol {u}^*$. The construction of the dataset and the design of the loss function are discussed in Section \ref{DNNdesign}.

\begin{remark}[Suboptimal Solution]
The proposed framework also applies to suboptimal feasible solutions, provided that the recursive feasibility and stability conditions of the suboptimal NMPC are preserved (see \cite{SuboptimalNMPC}) and the analysis in Section~\ref{stabilityanalysis} also applies to suboptimal solutions. For notational simplicity, the following analysis focuses on optimal solutions.
\end{remark}
\begin{remark}[Fitting $x\rightarrow {u}^*_{0|k}$]
Another approach is to fit the first control action $x\rightarrow {u}^*_{0|k}$, where $k\geq0$ is an arbitrary time step. Actually, $x\rightarrow {u}^*_{0|k}$ is a special case of $x\rightarrow \boldsymbol {u}^*$ and fitting the latter offers a more general and intuitive basis for the subsequent stability analysis.
\end{remark}

\indent Replacing the optimization process with a well-trained DNN yields the approximate NMPC. 
\subsection{The Approximation Error}
\label{approximationerror}
Since perfect approximation by the network is unattainable, the approximation error is introduced.

Within the feasible region of the optimization problem, the approximation error sequence is defined as
\[\boldsymbol{e}^{\mathrm{nn}}_k := \boldsymbol{u}_k ^{\mathrm{nn}}-\boldsymbol {u}_k ^*=\{e^{\mathrm{nn}}_{0|k},e^{\mathrm{nn}}_{1|k},...,e^{\mathrm{nn}}_{N_p-1|k}\},\quad k\geq0.\]
\indent As shown in Section \ref{section:DNN}, a neural network with continuous activation functions is a continuous function. Therefore, its output is bounded for bounded inputs. Since both $\mathcal{U}$ and $\mathcal{X}$ are compact, $\boldsymbol{u}^*_k$ and $\boldsymbol {u}^{\mathrm{nn}}_k$ are bounded and have well-defined extrema. Consequently, there exists a maximum $\overline{e}$ on the approximation error such that $\|e^{\mathrm{nn}}_{i|k}\|\leq \overline{e}$ for $i\in\mathbb{I}_{0:N_p-1}$.

\subsection{The General Idea}
The proposed framework follows an iterative trial-and-error procedure. 

First, the NMPC constraints are tightened to account for a prescribed maximum approximation error. The resulting tightened NMPC is then used to generate training data for the approximate NMPC. After training, a verification scheme estimates the probability that the approximation error satisfies the prescribed bound throughout the entire closed-loop trajectory. If the requirement is not met, the dataset is enlarged and the DNN is retrained. 

\section{Learning the NMPC}
\label{section:learning}
This section presents the approximate NMPC framework. The constraints of Problem \ref{prob1} are tightened to address the approximation error, as detailed in Section \ref{NMPCdesign}. Section \ref{DNNdesign} describes the construction of the training dataset and the design of the DNN. Finally, Section \ref{stabilityanalysis} establishes the constraint satisfaction and closed-loop stability under the approximate NMPC.

\subsection{NMPC with Tightened Constraints}
\label{NMPCdesign}

Tightening Problem \ref{prob1} into Problem \ref{prob2}, which is defined as follows:
\begin{prob}\label{prob2}
    \begin{align*}
    &\mathop{\min}\limits_{\boldsymbol {u}_k}  \ J(x_k,\boldsymbol {u}_k )\\
s.t. \quad & \overline{x}_{0|k}=x_k,\quad \overline{x}_{i+1|k}=f(\overline{x}_{i|k},u_{i|k}), \\
& u_{i|k} \in \mathcal{U}',\quad \overline{x}_{i|k} \in \mathcal{X}_i',\quad i \in\mathbb{I}_{0:N_p-1},\\
& \overline{x}_{N_p|k} \in \mathcal{X}'_f,\
\end{align*}
\end{prob}
where $\mathcal{U}'\subset \mathcal{U}$, $\mathcal{X}'_i \subset \mathcal{X}$ and $\mathcal{X}_f'\subset\mathcal{X}_f$.

This subsection presents the tightening strategies for control, state, and terminal constraints, and ultimately constructs the tightened NMPC for dataset generation. 

\subsubsection{Control Constraint Tightening}
\begin{theorem}[Control Constraint Tightening]
\label{tightenU}
Suppose that $\mathcal{U'}\oplus\mathcal{B}(\overline e)\subseteq \mathcal{U}$, then $u+e^{\mathrm{nn}} \in \mathcal{U}$ holds for any $u\in\mathcal{U}'$ and $e^{\mathrm{nn}}$ with $\|e^{\mathrm{nn}}\|\leq \overline{e}$. 
\end{theorem}
\begin{proof}
Since $\|e^{\mathrm{nn}}\|\leq\overline e$, we have $e^{\mathrm{nn}}\in\mathcal{B}(\overline e)$. Then
\[
u+e^{\mathrm{nn}}
\in\mathcal{U}'\oplus\mathcal{B}(\overline e)
\subseteq\mathcal{U}.\myqedhere
\]
\end{proof}

\indent If $\mathcal{U}$ is convex, $\mathcal{U}'$ can be obtained by linear scaling:
\[\mathcal{U}' = \beta_1 \mathcal{U}\subset\mathcal{U},\quad\beta_1\in(0,1).\]
\indent Then a sufficient condition is 
\[(1-\beta_1)r_{\min} \geq \overline{e},\]
where $r_{\min}:=\mathop{\min}\limits_{r\in \overline{\mathcal{U}}}\|r\|$ and $\overline{\mathcal{U}}$ is the boundary of $\mathcal{U}$.
\subsubsection{State Constraint Tightening}
In the presence of control errors, the predicted trajectory of system (\ref{sys}) from the same initial state can be divided into: (i) the nominal predicted trajectory $\overline{\boldsymbol{x}}_k$ under the optimal control sequence $\boldsymbol{u}^*_k$; and (ii) the approximate predicted trajectory $\boldsymbol{x}^{\mathrm{nn}}_k$ under the approximate control sequence $\boldsymbol{u}^{\mathrm{nn}}_k$.

Define the state deviation trajectory as 
\[\mathit{\delta} \boldsymbol{x}_k:=\boldsymbol{x}^{\mathrm{nn}}_k-\overline{\boldsymbol{x}}_k=\{\mathit{\delta} x_{0|k},\mathit{\delta} x_{1|k},...,\mathit{\delta} x_{N_p|k}\}.\]
Since the approximation error is bounded, the resulting state deviation $\mathit{\delta} x_{i|k}$ is also bounded.

For system (\ref{sys}), since $f$ is continuous differentiable on $\mathcal{X} \times \mathcal{U}$ where $\mathcal{X},\mathcal{U}$ are compact sets, its partial derivatives $\|\frac{\partial f}{\partial x} \|$ and $\|\frac{\partial f}{\partial u} \|$ are bounded on $\mathcal{X} \times \mathcal{U}$. Consequently, $f$ is Lipschitz continuous with respect to both $x,u$ on $\mathcal{X} \times \mathcal{U}$ with the Lipschitz constants $v_1,v_2 \ge 0$,
\begin{equation}\label{lipschitz}
\begin{aligned}&\|f(x_1,u)-f(x_2,u)\| \le v_1\|x_1-x_2 \|,\\
&\|f(x,u_1)-f(x,u_2)\| \le v_2\|u_1-u_2\|.\end{aligned}
\end{equation}
\begin{lemma}[State Deviation]
\label{statedeviation}
Suppose that, for any $k\geq0$, $\overline{x}_{i|k},x^{\mathrm{nn}}_{i|k}\in\mathcal{X}$ for $i\in \mathbb{I}_{0:N_p}$ and $u^*_{j|k}\in\mathcal{U}'$ with $\|e^{\mathrm{nn}}_{j|k}\|\leq\bar e$ for $j\in \mathbb{I}_{0:N_p-1}$. Then the state deviation trajectory $\mathit{\delta} \boldsymbol{x}_k$~satisfies
\begin{equation}\label{state_devation}
    \begin{aligned}\|\mathit{\delta}x_{i|k}\|\leq \overline{\delta}_{i}:= \left \{ \begin{aligned}  &0 &&, i = 0,\\
     &  \sum_{j=0}^{i-1}v_1^jv_2\overline{e}&&, i\in\mathbb{I}_{1:N_p}.\end{aligned} \right.
\end{aligned}
\end{equation}
\end{lemma}
\begin{proof}
Since $x^{\mathrm{nn}}_{0|k}=\bar{x}_{0|k}$, we have $\delta x_{0|k}=0$. For $i\in\mathbb{I}_{1:N_p}$, applying inequality (\ref{lipschitz}) recursively yields,
\begin{align*}
\|\delta x_{i|k}\|
&\leq v_1\|\delta x_{i-1|k}\|
   +v_2\overline{e} \\
&\leq
\sum_{j=0}^{i-1}v_1^jv_2\overline{e}.
\myqedhere
\end{align*}
\end{proof}
\begin{theorem}[State Constraint Tightening]
\label{tightenX}
For any $k\geq0$, suppose that $u^*_{i|k}\in\mathcal{U}'$ with $\|e^{\mathrm{nn}}_{i|k}\|\leq\bar e$, and
\begin{align*}
&\mathcal{X}'_i\oplus\mathcal{B}(\overline{\delta}_i)
\subseteq\mathcal{X},
&&i\in\mathbb{I}_{0:N_p-1},\\
&\mathcal{X}'_{j+1}\oplus
\mathcal{B}(v_1^jv_2\overline{e})
\subseteq\mathcal{X}'_j,
&&j\in\mathbb{I}_{0:N_p-2}.
\end{align*}
\indent Then $x^{\mathrm{nn}}_{i|k}=\overline{x}_{i|k}+\delta x_{i|k}\in\mathcal{X}$ holds for any $\overline{x}_{i|k}\in\mathcal{X}'_i$.
\end{theorem}
\begin{proof}
By Theorem \ref{tightenU}, $u^{\mathrm{nn}}_{i|k}\in\mathcal{U}$ for all $i\in\mathbb{I}_{0:N_p-1}$. Since $\delta x_{0|k}=0$, it follows that $x^{\mathrm{nn}}_{0|k}=\overline{x}_{0|k}
\in\mathcal{X}'_0\subseteq\mathcal{X}$. 

By inequality (\ref{lipschitz}), $\delta x_{1|k}$ satisfies $\|\delta x_{1|k}\|\leq v_2\overline{e}=  \overline{\delta}_{1}$. Hence, $\delta x_{1|k}\in\mathcal{B}(\overline{\delta}_1)$ and
\[
x^{\mathrm{nn}}_{1|k}=\overline{x}_{1|k}+\delta x_{1|k}\in\mathcal{X}'_1\oplus\mathcal{B}(\overline{\delta}_1)
\subseteq\mathcal{X}.
\]
\indent Repeating the proof above yields $x^{\mathrm{nn}}_{i|k}=\overline{x}_{i|k}+\delta x_{i|k}\in\mathcal{X}$ for any $\overline{x}_{i|k}\in\mathcal{X}'_i$.
\end{proof}

\indent If $\mathcal{X}$ is convex, $\mathcal{X}_i'$ can be obtained by linear scaling:
\[\mathcal{X}_i' = \beta_{i|2} \mathcal{X},\quad \beta_{i|2}\in(0,1],\quad i\in\mathbb{I}_{0:N_p-1}.\]
\indent Then a sufficient condition is
\begin{align*}
    &(1-\beta_{i|2})r_{\min} \geq \overline{\delta}_{i}, &&i\in\mathbb{I}_{0:N_p-1},\\
&(\beta_{j|2}-\beta_{j+1|2})r_{\min} \geq  v_1^{j}v_2\overline{e},&&j\in\mathbb{I}_{0:N_p-2},
\end{align*}
where $r_{\min}:=\mathop{\min}\limits_{r\in \overline{\mathcal{X}}}\|r\|$ and $\overline{\mathcal{X}}$ is the boundary of $\mathcal{X}$.

\subsubsection{Terminal Constraint Tightening}

Since $\mathcal{X}_f$ is convex, $\mathcal{X}'_f $ can be obtained by scaling $\alpha$:
\[\mathcal{X}'_f := \{x \in \mathbb{R}^{n_x} | x^\top  Px \leq \beta_3\alpha\},\quad \beta_3\in (0,1).\]
\indent For $\overline{x}_{N_p|k}\in\mathcal{X}'_f$ and $\|\delta x_{N_p|k}\|\leq\overline{\delta}_{N_p}$, the deviated terminal state satisfies
\[x^{\mathrm{nn}}_{N_p|k}=\overline{x}_{N_p|k}+\mathit{\delta}x_{N_p|k}\in\tilde{\mathcal{X}}'_f:=\mathcal{X}'_f\oplus\mathcal{B}(\overline{\delta}_{N_p}).\]

\begin{lemma}[The Form of $\tilde{\mathcal{X}}'_f$]
\label{form_tightenX_f} 
\begin{align*}
    \tilde{\mathcal{X}}'_f\subseteq\Big\{x\in \mathbb{R}^{n_x}|x^\top Px\leq\kappa_f^2\Big\},
\end{align*}
where $\kappa_f:=\sqrt{\lambda_{\max}(P)}\overline{\delta}_{N_p}
+\sqrt{\beta_3 \alpha}$.
\end{lemma}

\begin{proof}
Let $\overline{x}=\overline{x}_{N_p|k}$, $d=\mathit{\delta}x_{N_p|k}$ and $x=\overline{x}+d$, then it follows that
\[
x^\top Px
=\overline{x}^\top P\overline{x}+2\overline{x}^\top Pd+d^\top Pd,
\]
where $\overline{x}^\top P\overline{x}\leq\beta_3\alpha$ and $d^\top Pd\leq\lambda_{\max}(P)\overline{\delta}_{N_p}^2$.

By the Cauchy-Schwarz Inequality,
\begin{align*}
    \overline{x}^\top Pd&\leq\sqrt{\overline{x}^\top P\overline{x}}\sqrt{d^\top Pd}\leq\sqrt{\lambda_{\max }(P)}\sqrt{\beta_3 \alpha}\overline{\delta}_{N_p}.
\end{align*}
\indent Therefore,
\[
    x^\top Px\leq\left(\sqrt{\lambda_{\max }(P)}\overline{\delta}_{N_p}+\sqrt{\beta_3 \alpha}\right)^2.\myqedhere
\]
\end{proof}

\begin{theorem}[Terminal Constraint Tightening]
\label{tightenX_f}
Suppose $\tilde{\mathcal{X}}'_f$ satisfies the following properties:

1) $\tilde{\mathcal{X}}'_f\subseteq \mathcal{X}_f$;

2) Under the terminal feedback control law $K$, $x\in\tilde{\mathcal{X}}'_f$ returns to $\mathcal{X}'_f$ at the next step.

\indent
A sufficient condition is that the contraction factor $\beta_3\in (0,1)$ satisfies
\begin{align*}
\frac{\sqrt{\rho}}{1-\sqrt{\rho}}\nu &\leq\sqrt{\beta_3}\leq 1-\nu ,\quad
\nu\leq 1-\sqrt{\rho},
\end{align*}
where $\rho := 1-\frac{\lambda_{\min }\left(Q_K\right)}{\lambda_{\max }(P)}$ and $\nu:=\sqrt{\frac{\lambda_{\max }(P)}{\alpha}}\overline{\delta}_{N_p}$.
\end{theorem}
\begin{proof}From Lemma \ref{form_tightenX_f}, to satisfy the property 1), it is required that 
\begin{equation}\label{inequ1}
    \left(\sqrt{\lambda_{\max }(P)}\overline{\delta}_{N_p}+\sqrt{\beta_3 \alpha}\right)^2\leq \alpha.
\end{equation}
\indent By Lemma \ref{terminalcontrollaw}, $u = Kx$ guarantees, for $x\in \mathcal{X}_f$, 
\[V_f(x^+)-V_f(x)\leq -x^\top Q_Kx,\]
then, it holds that
\[V_f(x^+)\leq \left(1-\frac{\lambda_{\min }\left(Q_K\right)}{\lambda_{\max }(P)}\right)V_f(x)= \rho V_f(x).\]
\indent To satisfy the property 2), it is required that 
\begin{equation}\label{inequ2}
    \rho\left(\sqrt{\lambda_{\max }(P)}\overline{\delta}_{N_p}+\sqrt{\beta_3 \alpha}\right)^2\leq \beta_3 \alpha.
\end{equation}
\indent Solving inequalities (\ref{inequ1}) and (\ref{inequ2}) yields
\begin{equation}\label{inequ3}
    \frac{\sqrt{\rho}}{1-\sqrt{\rho}}\nu \leq\sqrt{\beta_3}\leq 1-\nu.
\end{equation}
\indent Consider the existence condition of inequality~(\ref{inequ3}).

From equation (\ref{equ::terminalcontrollaw}), since $A^\top _KPA_K\succeq0$, it follows that $P \succ Q_K$, $\frac{\lambda_{\min }\left(Q_K\right)}{\lambda_{\max }(P)}<1$ and hence $\rho\in(0,1)$. Therefore, the term $\frac{\sqrt{\rho}}{1-\sqrt{\rho}}$ is well-defined.

To ensure that inequality (\ref{inequ3}) holds, it is required that 
\[
\nu \leq 1-\sqrt{\rho}.\myqedhere
\]
\end{proof}

\begin{remark}
The factor $\beta_3$ may require further adjustment to satisfy $\tilde{\mathcal{X}}'_f\subseteq\mathcal{X}_{N_p-1}'$ and $u=Kx\in\mathcal{U}'$ for $x\in \mathcal{\tilde{X}}'_f$ to ensure the feasibility of system (\ref{sys}) under the tightened NMPC. In the following, these two conditions are assumed to hold by default. 
\end{remark}

\subsection{Training the Approximate NMPC}
\label{DNNdesign}
\subsubsection{Sampling Strategy}
The dataset 
\[\Xi=\{(x_0,\boldsymbol{u}^*_0),(x_1,\boldsymbol{u}^*_1),...,(x_{D-1},\boldsymbol{u}^*_{D-1})\},\]
is constructed by pairs of initial state and the corresponding optimal (or suboptimal yet feasible) control sequence of Problem \ref{prob2}, where $D$ denotes the total number of samples.

Random sampling is performed over the state constraint set $\mathcal{X}$ to construct $\Xi$, where samples are selected independently and uniformly. If the sampled state is infeasible for Problem \ref{prob2}, it will be skipped. Therefore, $\Xi$ consists of i.i.d. samples drawn from the uniform distribution over the feasible region of Problem \ref{prob2}.

\subsubsection{Design of DNN}
The DNN is designed as a fully-connected architecture to fit the mapping $x\rightarrow \boldsymbol {u}^*$. The loss function $F_{\mathrm{loss}}$ is inspired by the cost of NMPC. 

To reduce the influence of large fitting residuals, the Cauchy robust loss~\cite{CauchyRobustloss} is adopted:
\[\rho_c(e,\omega):=\log(1+\frac{e^2}{\omega^2}),\quad \omega>0,\]
where $\omega$ is a scale parameter.

The loss function is defined as
\begin{align}
&F_{\mathrm{loss}}
(\boldsymbol{u}^{\mathrm{nn}}_k,\boldsymbol{u}^*_k,x_k)
\nonumber=
\omega^2(\lambda_1 L_{u,k}+\lambda_2 L_{x,k}+\lambda_3 L_{f,k})
,
\end{align}
where
\begin{align*}
L_{u,k}
&=
\frac{1}{2N_p}
\sum_{j=0}^{N_p-1}\rho_c\left(\|e^{\mathrm{nn}}_{j|k}\|,\omega\right)
\nonumber\\
&\quad+\frac{1}{2}
\rho_c\left(\max_{\substack{0\leq i\leq N_p-1}}
\|e^{\mathrm{nn}}_{i|k}\|,\omega\right),\\
L_{x,k}
&=
\frac{1}{N_p+1}
\sum_{i=0}^{N_p}
\rho_c\left(\|\delta x_{i|k}\|,\frac{\omega}{4}\right),\\
L_{f,k}
&=
\rho_c\left(
\sqrt{V_f(\delta x_{N_p|k})},2\omega
\right),\quad k\geq0,
\end{align*}
and $\lambda_1,\lambda_2,\lambda_3>0$ are weighting coefficients.

The control term combines the mean and worst-component penalties with equal weights. The latter two terms of the loss function expand the representation of error, which may enable the network to capture the dynamics of the entire trajectory.

%


Replacing the NMPC optimization process with a well-trained DNN yields the approximate NMPC. 

\subsection{Constraint Satisfaction and Stability Analysis}
\label{stabilityanalysis}
This section analyzes the constraint satisfaction along the closed-loop trajectory and the stability of the approximate NMPC.

\subsubsection{Recursive Feasibility under Approximation Error}

Although the approximate NMPC does not solve any optimization problem online, recursive feasibility of Problem~\ref{prob2} in the presence of approximation errors is essential to ensure that the target mapping $x\rightarrow\boldsymbol{u}^*$ remains well-defined along the closed-loop trajectory.

\begin{lemma}[Recursive Feasibility]
\label{feasibility}
Suppose $x_k$ is feasible for Problem~\ref{prob2} and $\|e^{\mathrm{nn}}_{i|k}\|\leq \overline{e}$, then $x_{k+1}$ remains feasible for Problem~\ref{prob2} for all $k\geq0$ and $i\in\mathbb{I}_{0:N_p-1}$.
\end{lemma}

\begin{figure}[!tbp]
\centering
\includegraphics[width=0.95\linewidth]{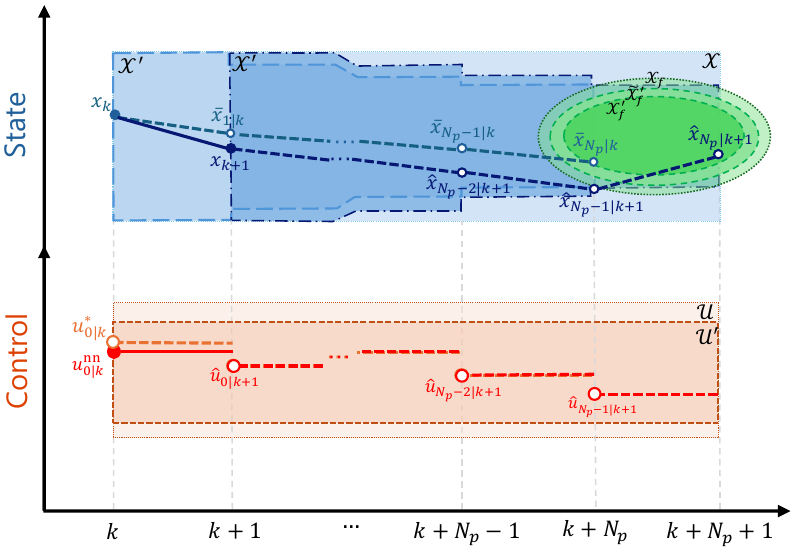}
\caption{Illustration of the predicted trajectories and control sequences at steps $k$ and $k+1$. Solid lines and dots denote actual state evolution and inputs, while dashed lines and hollow dots denote predicted ones.}
\label{fig:Proof}

\end{figure}

\begin{proof}
Since $x_k$ is feasible for Problem~\ref{prob2}, we have $\overline{x}_{i|k} \in \mathcal{X}'_i$ and $u^*_{i|k}\in \mathcal{U}'$ for $i\in\mathbb{I}_{0:N_p-1}$, and $\overline{x}_{N_p|k} \in \mathcal{X}_f'$.

At the next closed-loop state $x_{k+1}=f(x_k,u^{\mathrm{nn}}_{0|k})$, consider the shifted control sequence 
\begin{equation}\label{shiftedcontrol}
    \hat{\boldsymbol{u}}_{k+1}(x_{k+1}):=[u^*_{1|k},u^*_{2|k},\cdots,u^*_{N_p-1|k},Kz_{N_p-1}(x_{k+1})],
\end{equation}
where $\hat{u}_{N_p-1|k+1}(x_{k+1})$ is the terminal control law (see Lemma \ref{terminalcontrollaw}) and $z_{N_p-1}(x_{k+1})$ denotes the $(N_p-1)$-th predicted state from $x_{k+1}$ under $[u^*_{1|k},\cdots,u^*_{N_p-1|k}]$.

Define the trajectory deviation $ \mathit{\Delta}\boldsymbol{x}_{k+1}$ that satisfies
\[\mathit{\Delta}x_{i|k+1}=\hat{x}_{i|k+1}-\hat{x}^{\mathrm{nom}}_{i|k+1},\quad i\in\mathbb{I}_{0:N_p},\]
where $\hat{\boldsymbol{x}}_{k+1}$ and $\hat{\boldsymbol{x}}^{\mathrm{nom}}_{k+1}$ denote the trajectories starting from $x_{k+1}$ and $\overline{x}_{1|k}$ under $\hat{\boldsymbol{u}}_{k+1}(\cdot)$, respectively. The comparison of $\hat{\boldsymbol{x}}_{k+1}$ and $\hat{\boldsymbol{x}}^{\mathrm{nom}}_{k+1}$ is shown in Figure \ref{fig:PredictionTraj}.
\begin{figure}[!tbp]
\centering
\includegraphics[width=0.8\linewidth]{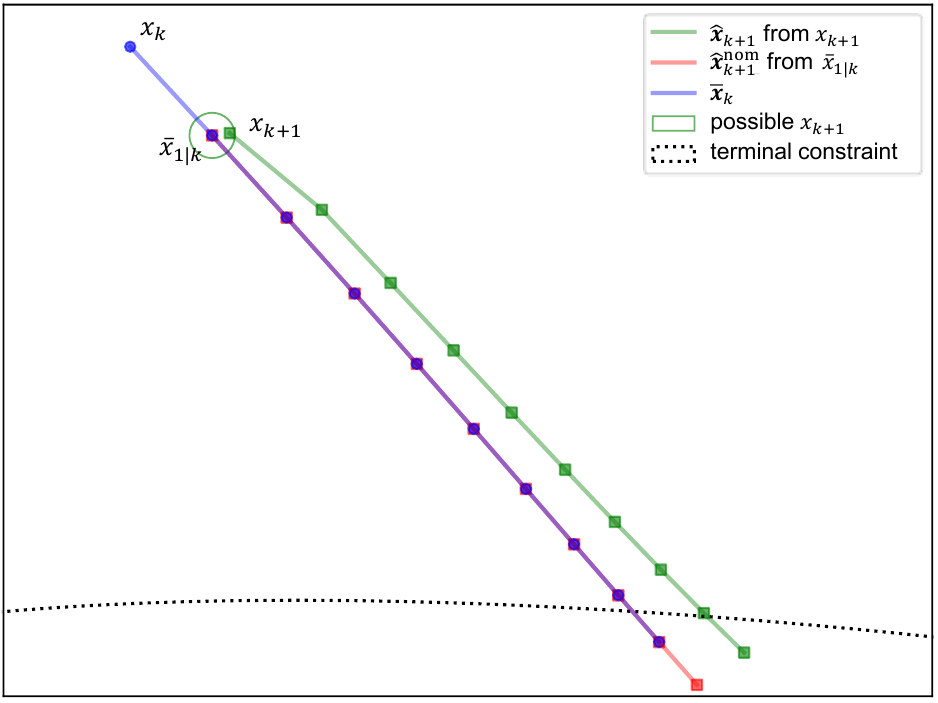}
\caption{Comparison of trajectories  $\hat{\boldsymbol{x}}_{k+1}$ and $\hat{\boldsymbol{x}}^{\mathrm{nom}}_{k+1}$. Due to $e^{\mathrm{nn}}_{0|k}$, the system state at step $k+1$ deviates from $\overline{x}_{1|k}$ to $x_{k+1}$, which consequently leads to a deviation in the predicted trajectories.}
\label{fig:PredictionTraj}

\end{figure}

The Lipschitz property of system (\ref{sys}) implies
\begin{equation}\label{Deva}
    \begin{aligned}
    \|\mathit{\Delta}x_{i|k+1}\|\leq \mathit{\overline{\Delta}}_{i}:= v_1^{i} v_2 \overline{e},\quad i\in\mathbb{I}_{0:N_p-1}.
\end{aligned}
\end{equation}
\indent By inequality (\ref{state_devation}), the state deviation satisfies
\[  \overline{\delta}_{i}:=\sum_{j=0}^{i-1} v_1^{j} v_2\overline{e}=\sum_{j=0}^{i-1}\mathit{\overline{\Delta}}_{j},\quad i\in \mathbb{I}_{1:N_p}.\]
\indent Together with Theorem \ref{tightenX}, the tightened state constraints satisfy 
\[\mathcal{X}'_{i+1}\oplus\mathcal{B}( \|\mathit{\Delta}x_{i|k+1}\|)\subseteq \mathcal{X}'_{i},\quad i\in\mathbb{I}_{0:N_p-2}.\]
\indent Consequently, since $\hat{x}^{\mathrm{nom}}_{i|k+1}=\overline{x}_{i+1|k}\in \mathcal{X}'_{i+1}$, the states along the predicted trajectory satisfy
\[
    \hat{x}_{i|k+1}=\hat{x}^{\mathrm{nom}}_{i|k+1}+\mathit{\Delta}x_{i|k+1}\in \mathcal{X}'_i,\quad i\in\mathbb{I}_{0:N_p-2}.
\]
\indent According to Theorem \ref{tightenX_f}, since $\hat{x}^{\mathrm{nom}}_{N_p-1|k+1}=\overline{x}_{N_p|k}\in \mathcal{X}'_f$, it follows that
\begin{align*}
 &\hat{x}_{N_p-1|k+1}\in  \tilde{\mathcal{X}}'_f\subseteq\mathcal{X}'_{N_p-1},\\
   & \hat{u}_{N_p-1|k+1}=K \hat{x}_{N_p-1|k+1}\in \mathcal{U}',\\
   & \hat{x}_{N_p|k+1}\in \mathcal{X}'_f.
\end{align*}
\indent Therefore, a feasible solution $\boldsymbol{\hat{u}}_{k+1}(x_{k+1})$ for Problem \ref{prob2} exists at $x_{k+1}$ and $x_{k+1}$ is feasible for Problem \ref{prob2}. The overall idea is illustrated in Figure \ref{fig:Proof}. 
\end{proof}
By Lemma~\ref{feasibility}, the closed-loop trajectory remains feasible for Problem~\ref{prob2} under the approximate NMPC. Hence, the corresponding solution $\boldsymbol{u}_k^*$ and approximation error $\boldsymbol{e}^{\mathrm{nn}}_k$ exist and satisfy
\[\boldsymbol {u}_k^{\mathrm{nn}}= \boldsymbol{e}^{\mathrm{nn}}_k +\boldsymbol{u}_k^*,\quad  k\geq0.\]

\subsubsection{Constraint Satisfaction}

\begin{theorem}[Constraint Satisfaction]
Consider the closed-loop system \eqref{sys} under the approximate NMPC. Suppose $x_k$ is feasible for Problem \ref{prob2} and $\|e^{\mathrm{nn}}_{i|k}\|\leq \overline{e}$ for $i\in\mathbb{I}_{0:N_p-1}$, then, for all $k\geq0$,
\[x_k,x_{k+1}\in\mathcal{X},\quad u_k=u_{0|k}^{\mathrm{nn}}\in\mathcal{U},\quad u_{k+1}=u_{0|k+1}^{\mathrm{nn}}\in\mathcal{U}.\]
\end{theorem}
\begin{proof}
Since $x_k$ is feasible for Problem \ref{prob2}, we have $x_k\in\mathcal{X}$ and $u^*_{0|k}\in\mathcal{U}'$.

By Theorem~\ref{tightenU}, since $\|e^{\mathrm{nn}}_{i|k}\|\leq \overline{e}$,
\[
u_k=u_{0|k}^{\mathrm{nn}}
=u^*_{0|k}+e_{0|k}^{\mathrm{nn}}
\in\mathcal{U}'\oplus\mathcal{B}(\overline e)
\subseteq\mathcal{U}.
\]
\indent By Lemma~\ref{feasibility}, Problem~\ref{prob2} remains feasible at $x_{k+1}$ for all $k\geq0$. Therefore,
\[
x_{k+1}\in\mathcal{X} ,\quad u_{k+1}=u_{0|k+1}^{\mathrm{nn}}\in\mathcal{U}.\myqedhere
\]
\end{proof}
\subsubsection{Stability}
Approximation errors may induce deviations in both control inputs and state trajectories, thereby affecting closed-loop stability. Inspired by \cite{YU20142269}, this section establishes conditions for ensuring closed-loop stability from the perspective of suboptimal solutions. 

Our goal is to analyze how approximation errors affect the value function, including the deviation in the control sequence (between $\boldsymbol{u}_k^*$ and $\boldsymbol{u}_k^{\mathrm{nn}}$) and the initial state (between $x_{k+1}$ and $\overline{x}_{1|k}$), and derive sufficient conditions for its monotonic decrease. 

Define the deviation sets
\begin{align*}
&\Psi(\boldsymbol{u}^*_k):=  \{\boldsymbol{u} \in \mathbb{R}^{N_p\times n_u } |\|u_{i|k}-u_{i|k}^*\|\leq\overline e,\ i\in\mathbb I_{0:N_p-1}\},\\
&\Phi(\overline{x}_{1|k}):=  \{x \in \mathbb{R}^{n_x} | \|x-\overline{x}_{1|k}\| \leq \overline{\delta}_{1}\}.
\end{align*}
\indent For the same feasible initial state $x_k$, consider the deviation in the cost function induced by using $\boldsymbol{u}_k^*$ and $\boldsymbol{u}^{\mathrm{nn}}_k$. 

For $\boldsymbol{u}^{\mathrm{nn}}_k\in\Psi(\boldsymbol{u}_k^*)$, by Assumption \ref{assu:2}, $J(x_k,\boldsymbol{u}^{\mathrm{nn}}_k)$ is continuous with respect to $\boldsymbol{u}^{\mathrm{nn}}_k$. Since $J(x_k,\boldsymbol{u}^{\mathrm{nn}}_k)\geq 0$, there exists a lower bound. Since $\Psi(\boldsymbol{u}_k^*)$ is compact, $J(x_k,\boldsymbol{u}^{\mathrm{nn}}_k)$ has an upper bound.

For the same feasible initial state $x_k$, let
\[\mathit{\Delta}\mathit{M}(\boldsymbol{u}_k^*):=\max\limits_{\boldsymbol{u}^{\mathrm{nn}}_k}\left(|J(x_k,\boldsymbol{u}^{\mathrm{nn}}_k)-J(x_k,\boldsymbol{u}_k^*)|\right)\]
denote the maximum cost deviation for $\boldsymbol{u}^{\mathrm{nn}}_k\in\Psi (\boldsymbol{u}_k^*)$.

\begin{lemma}[Control Sequence Cost Deviation]
\label{DeltaMto0}
Suppose $\|e^{\mathrm{nn}}_{i|k}\|\leq \overline{e}$ for $i\in\mathbb{I}_{0:N_p-1}$, then there exists a $\mathcal{K}_{\infty}$ function $\gamma_1(\cdot)$ such that, for all $\boldsymbol{u}_k^*$,
\[
\mathit{\Delta}\mathit{M}(\boldsymbol{u}_k^*)
\leq \mathit{\Delta}\mathit{\overline{M}}:= 
\gamma_1\!\left(\overline{e}\right).
\]
\end{lemma}
\begin{proof}
For $\boldsymbol{u}^{\mathrm{nn}}_k\in\Psi(\boldsymbol{u}_k^*)$, 
\begin{equation}\label{ControlSeq_Deva}
    \begin{aligned}
&\left|J(x_k,\boldsymbol{u}^{\mathrm{nn}}_k)-J(x_k,\boldsymbol{u}_k^*)\right|\\
&\leq\sum_{i=0}^{N_p-1}  \big(\left|\|x^{\mathrm{nn}}_{i|k}\|_{Q}^{2}-\|\overline{x}_{i|k}\|_{Q}^{2}\right|+\left|\|u^{\mathrm{nn}}_{i|k}\|^2_R-\|u^{*}_{i|k}\|^2_R\right|\big)\\&
\qquad+\left|\|x^{\mathrm{nn}}_{N_p|k} \|_{P}^{2}-\|\overline{x}_{N_p|k} \|_{P}^{2}\right|.
\end{aligned}
\end{equation}
For any vectors $p,q$ and matrix $W\succeq0$, the following inequality holds 
\begin{equation}\label{inequ_cost_deva}
    \bigl|\|p+q\|_W^2-\|p\|_W^2\bigr|\leq 2\|p\|_W\|q\|_W+\|q\|_W^2.
\end{equation}
Since 
\begin{align*}
    &x^{\mathrm{nn}}_{i|k}=\overline{x}_{i|k}+\mathit{\delta}x_{i|k},\quad &&i\in\mathbb{I}_{0:N_p},\\
    &u_{j|k}^{\mathrm{nn}}=u_{j|k}^*+e_{j|k}^{\mathrm{nn}},\quad &&j\in\mathbb{I}_{0:N_p-1},
\end{align*}
while $Q,R$ and $P$ are positive definite, inequality (\ref{inequ_cost_deva}) holds for inequality (\ref{ControlSeq_Deva}).

By inequality (\ref{state_devation}), $\|\mathit{\delta}x_{i|k}\|$ is linearly bounded in $\overline{e}$, while $\|e^{\mathrm{nn}}_{j|k}\|
\leq\overline{e}$. Since $\mathcal{X}'_j$, $\mathcal{X}'_f$ and $\mathcal{U}'$ are compact, the remaining terms are uniformly bounded by constants. 

Taking these terms at their upper bounds yields that, for $c_1,c_2>0$, 
\[
\mathit{\Delta}\mathit{M}(\boldsymbol{u}_k^*)
\leq
c_1\overline{e}
+c_2\overline{e}^2
=: \gamma_1\!\left(\overline{e}\right).
\]
\indent Since $\gamma_1(\cdot)$ is a $\mathcal{K}_{\infty}$ function, the result follows.
\end{proof}

\indent Consider the cost function deviation between $\boldsymbol{\hat{x}}_{k+1}$ and $\boldsymbol{\hat{x}}^{\mathrm{nom}}_{k+1}$, which originate from $x_{k+1}$ and $\overline{x}_{1|k}$ and are controlled by $\boldsymbol{\hat{u}}_{k+1}(\cdot)$, see inequality (\ref{shiftedcontrol}) and Lemma \ref{feasibility}.

Define
\[\hat{J}(x_{k+1}):=J(x_{k+1},\boldsymbol{\hat{u}}_{k+1}(x_{k+1})).\]
\indent The continuity of $f$ and $\hat{u}_{N_p-1|k+1}(x_{k+1})$ implies the continuity of $\hat{J}(x_{k+1})$ for $x_{k+1}\in \Phi(\overline{x}_{1|k})$. Then since $\Phi(\overline{x}_{1|k})$ is compact, $\hat{J}(x_{k+1})$ has an upper bound. Since $\hat{J}(x_{k+1})\geq 0$, there exists a lower bound.

Define 
\begin{align*}
\mathit{\Delta}&\mathit{N}(\overline{x}_{1|k}):=\max\limits_{x_{k+1}}\big(\big|\hat{J}(x_{k+1})-\hat{J}(\overline{x}_{1|k})\big|\big)
\end{align*}
as the maximum cost deviation for $x_{k+1}\in\Phi(\overline{x}_{1|k})$.

\begin{lemma}[Initial State Cost Deviation]
\label{DeltaNto0}
There exists a $\mathcal{K}_{\infty}$ function $\gamma_2(\cdot)$ such that, for all $\overline{x}_{1|k}$,
\[
\mathit{\Delta}\mathit{N}(\overline{x}_{1|k})
\leq\mathit{\Delta}\mathit{\overline{N}}:= \gamma_2\!\left(\overline{e}\right).
\]
\end{lemma}
\begin{proof}
For $x_{k+1}\in\Phi(\overline{x}_{1|k})$,
\begin{equation}\label{InitialState_Deva}
    \begin{aligned}
\big|&\hat{J}(x_{k+1})-\hat{J}(\overline{x}_{1|k})\big|\\
&\leq\sum_{i=0}^{N_p-1}  \big(\left|\|\hat{x}_{i|k+1}\|_{Q}^{2}-\|\hat{x}^{\mathrm{nom}}_{i|k+1}\|_{Q}^{2}\right|\big)\\
&\quad+ \left|\|\hat{x}_{N_p|k+1} \|_{P}^{2}-\|\hat{x}^{\mathrm{nom}}_{N_p|k+1} \|_{P}^{2}\right|\\
&\quad+ \left|\|K\hat{x}_{N_p-1|k+1}\|^2_R-\|K\hat{x}^{\mathrm{nom}}_{N_p-1|k+1}\|^2_R\right|.
\end{aligned}
\end{equation}
\indent Since $\hat{x}_{i|k+1}=\hat{x}^{\mathrm{nom}}_{i|k+1}+\mathit{\Delta}x_{i|k+1}$ for $i\in\mathbb{I}_{0:N_p}$, inequality (\ref{inequ_cost_deva}) holds for inequality (\ref{InitialState_Deva}).

Following inequality (\ref{Deva}), $\|\mathit{\Delta}x_{j|k+1}\|\leq v_1^{j} v_2 \overline{e}$ for $j\in\mathbb{I}_{0:N_p-1}$. By the Lipschitz property of system (\ref{sys}),
\begin{align*}
    \|\mathit{\Delta}x_{N_p|k+1}\|&\leq v_1 \|\mathit{\Delta} x_{N_p-1|k+1}\|
   +v_2\|K\mathit{\Delta} x_{N_p-1|k+1}\|\\
   & \leq(v_1+v_2\|K\|) v_1^{N_p-1} v_2 \overline{e}.
\end{align*}
Since $\mathcal{X}'_i$ and $\mathcal{X}'_f$ are compact, the remaining terms are uniformly~bounded.

Taking these terms at their upper bounds yields that, for $d_1,d_2>0$, 
\[
\mathit{\Delta}\mathit{N}(\overline{x}_{1|k})
\leq
d_1\overline{e}
+d_2\overline{e}^2
=: \gamma_2\!\left(\overline{e}\right).
\]
\indent Since $\gamma_2(\cdot)$ is a $\mathcal{K}_{\infty}$ function, the result follows.
\end{proof}

\begin{theorem}[Ultimately Bounded]
\label{ISS}
Suppose $x_k$ is feasible for Problem \ref{prob2} and $\|e^{\mathrm{nn}}_{i|k}\|\leq \overline{e}$ for $i\in\mathbb{I}_{0:N_p-1}$ and $k\geq0$, then the closed-loop system is ultimately bounded.
\end{theorem}
\begin{proof}
By Lemma \ref{terminalcontrollaw}, Lemma \ref{DeltaMto0} and Lemma \ref{DeltaNto0},
\begin{align*}
J(&x_{k+1},\boldsymbol{u}^{\mathrm{nn}}_{k+1})-J(x_{k},\boldsymbol{u}^{\mathrm{nn}}_k)\\
 &\leq J(x_{k+1},\boldsymbol{u}^*_{k+1})+\mathit{\Delta}\mathit{\overline{M}}-J(x_{k+1},\boldsymbol{\hat{u}}_{k+1})\\
 &\qquad+J(x_{k+1},\boldsymbol{\hat{u}}_{k+1})-J(x_{k},\boldsymbol{u}^*_{k})+\mathit{\Delta}\mathit{\overline{M}}\\
 &\leq 2\mathit{\Delta}\mathit{\overline{M}}+J(\overline{x}_{1|k},\boldsymbol{\hat{u}}_{k+1})-J(x_{k},\boldsymbol{u}^*_{k})+\mathit{\Delta}\mathit{\overline{N}}\\
 &\leq2\mathit{\Delta}\mathit{\overline{M}}+\mathit{\Delta}\mathit{\overline{N}} -x_{k}^\top Q x_{k}-u_{0|k}^{*,\top} Ru^*_{0|k}\\
 &\leq -\lambda_{\min}(Q)\|x_k\|^2+2\gamma_1(\overline{e})+\gamma_2(\overline{e}).
\end{align*}
\indent Therefore, the closed-loop system is uniformly ultimately bounded. 
\end{proof}
\begin{theorem}[Asymptotic Stability]
\label{AsymptoticStability}
For $i\in\mathbb{I}_{0:N_p-1}$ and $k\geq0$, suppose $x_k$ is feasible for Problem \ref{prob2}, $\|e^{\mathrm{nn}}_{i|k}\|\leq \overline{e}$, and\[
2\mathit{\Delta}\mathit{\overline{M}}+\mathit{\Delta}\mathit{\overline{N}}
<
\frac{\lambda_{\min}(Q)}
{\lambda_{\max}(P)}\alpha,
\]
then, under the dual-mode control law
\[
u(x_k)=
\begin{cases}
u^{\mathrm{nn}}_{0|k},&
x_k\in\mathcal{X}\setminus\mathcal{X}_f,\\
Kx_k,&
x_k\in\mathcal{X}_f,
\end{cases}
\]
the closed-loop system is asymptotically stable, where $\mathcal{X} \setminus\mathcal{X}_f:=\{x|x\in\mathcal{X}, x \notin \mathcal{X}_f\}$.
\end{theorem}
\begin{proof}
For $x_k\in\mathcal{X}\setminus\mathcal{X}_f$, $x_k^\top Px_k\geq\alpha$. Since $P,Q\succ0$,
\begin{align*}
    x_{k}^\top P x_{k}\leq \lambda_{\max}(P)\|x_{k}\|^2,\\ 
 x_{k}^\top Q x_{k}\geq \lambda_{\min}(Q)\|x_k\|^2.
\end{align*}
Therefore,
\[
\begin{aligned}
&J(x_{k+1},\boldsymbol{u}^{\mathrm{nn}}_{k+1})
-J(x_k,\boldsymbol{u}^{\mathrm{nn}}_k)
\\
&\leq
2\mathit{\Delta}\mathit{\overline{M}}+\mathit{\Delta}\mathit{\overline{N}}
-\frac{\lambda_{\min}(Q)}
{\lambda_{\max}(P)}\alpha
<0 .
\end{aligned}
\]
\indent Hence, the Lyapunov function strictly decreases outside $\mathcal{X}_f$, and the state enters $\mathcal{X}_f$ in finite time.

Once inside $\mathcal X_f$, the terminal feedback $u=Kx$ guarantees positive invariance and asymptotic stability by Lemma~\ref{terminalcontrollaw}. Therefore, the closed-loop system is asymptotically stable.
\end{proof}

\subsection{A Posteriori Verification of Approximation Error}
\label{estimateframework}
This section introduces a scheme to estimate a lower bound on the probability that the approximation error satisfies a prescribed maximum with a confidence level. Under the conditions established above, this probability bound also provides a lower bound on the probability of closed-loop constraint satisfaction and stability. The method of \cite{Hertneck2018} is reformulated using Empirical Bernstein Inequality (EBI) \cite{bernshteintheory,BernsteinInequality2} to reduce the conservatism of the verification.

For each sampled initial state $x_j$ whose trajectory reaches \(\mathcal{X}_f\), let 
\begin{align*}
\boldsymbol{u}^{*,(j)}(x_j)
&:=\{\{u_{i|k}^{*,(j)}\}_{i=0}^{N_p-1}\}_{k=0}^{T_j},\\
\boldsymbol{u}^{\mathrm{nn},(j)}(x_j)
&:=\{\{u_{i|k}^{\mathrm{nn},(j)}\}_{i=0}^{N_p-1}\}_{k=0}^{T_j},
\end{align*}
denote the optimal and approximate control sequence along the closed-loop trajectory, respectively, where \(T_j\) denotes the number of sampling steps required for the trajectory to enter \(\mathcal{X}_f\). Trajectories starting in \(\mathcal{X}_f\) are classified as safe, while those failing to reach it within the prescribed time limit or encountering optimization infeasibility are classified as unsafe.

Define the maximum approximation error along the closed-loop trajectory as
\[
\begin{aligned}
\mathcal{E}_{\mathrm{cl}}
(x_j)
:=\max_{k\in\mathbb{I}_{0:T_j-1}}\max_{i\in\mathbb{I}_{0:N_p-1}}
\|u_{i|k}^{\mathrm{nn},(j)}
-u_{i|k}^{*,(j)}\|.
\end{aligned}
\]
\indent Then define the safety indicator function as
\[I(x_j):=\left\{\begin{array}{ll}
1 ,& \text { if }\mathcal{E}_{\mathrm{cl}}
(x_j)
\leq\overline{e}_{\mathrm{pre}}, \\
0, & \text { otherwise.}
\end{array}\right.\]
\indent Since all constraints are closed sets and $f$ is continuous, the feasible region of Problem \ref{prob2} is also a closed set. 
Then define $P_X$ as a uniform distribution over the feasible region. The procedure randomly samples $p$ i.i.d. initial states $\{x_j\}^{p-1}_{j=0}$ from $P_X$ and computes the corresponding indicator set $\{I(x_j)\}^{p-1}_{j=0}$. Define the actual safety probability as $\mu := \mathbb{P}_{x\sim P_X}[I(x)=1]$ and the empirical safety probability as $\hat{\mu} :=\frac{1}{p}\sum_{j=0}^{p-1}I(x_j)$.

Note that the variance $\mathrm{Var}_{x\sim P_X}(I(x))$ can be relatively small for a well-fitted neural network. Therefore, EBI is employed to estimate $\mu$ from $\hat{\mu}$.

\begin{lemma}[Empirical Bernstein Inequality \cite{BernsteinInequality2}]
\label{Lemma::EBI}
For an i.i.d. set of states $\{x_j\}^{p-1}_{j=0}$ sampled from $P_X$, consider its indicator set $\{I(x_j)\}^{p-1}_{j=0}$ with empirical mean $\hat{\mu}$ and empirical variance $\hat{\sigma}^2$. By the EBI, with probability at least $1-\eta$, the following inequality holds:
\begin{equation}\label{EBI}
   \hat{\mu}- \mu\leq \sqrt{\frac{2 \hat{\sigma}^2 \ln \frac{2}{\eta}}{p}}+\frac{7 \ln\frac{2}{\eta}}{3(p-1)}=:\epsilon,
\end{equation}
where $\eta\in(0,1)$ is the significance level and $\epsilon$ is the confidence radius.
\end{lemma}

Given a desired significance level $\eta_d$ and define $\epsilon_d$ as its corresponding confidence radius. Then, with a confidence of at least $1-\eta_d $,
\[ \mathbb{P}_{x\sim P_X}[I(x)=1]=\mu\geq \hat{\mu}-\epsilon_d=\mu_{\mathrm{lower}}.\]
\begin{remark}
    In the original reference \cite{BernsteinInequality2}, the form of EBI is \[\mathbb{P}[\mu-\hat{\mu}\leq\epsilon]\geq 1-\eta.\]
\indent Noting that $I(x)$ and $1-I(x)$ share the same empirical variance, the original EBI can be reformulated into the form of inequality (\ref{EBI}). 
\end{remark}

\indent Since $P_X$ is fixed and the samples are i.i.d. drawn from $P_X$, as $p\rightarrow\infty$, it follows that $\epsilon_d\rightarrow0$, and by the law of large numbers, $\hat{\mu}$ converges almost surely to $\mu$. Consequently, $\mu_{\mathrm{lower}}$ converges almost surely to $\mu$.

The framework for verifying the approximation error is shown in Algorithm \ref{algorithm1}. If the verification fails, it should be restarted.
\begin{algorithm}[h]
    \caption{Approximation Error Verification}
    \label{algorithm1}
    \begin{algorithmic}[1]
        \Require $\overline{e}_{\mathrm{pre}}$, $\eta_d$, $p$, the desired safety probability $\mu_d$, expansion factor
        $n>1$, iteration limit $i_{\max}$.
        
        \State $i \gets 0$.
        
        \While{$i < i_{\max}$}
            \State Sample $p$ feasible initial states $\{x_j\}^{p-1}_{j=0}$.
            
            \State Compute $\{I(x_j)\}^{p-1}_{j=0}$, $\hat{\mu}$, and $\hat{\sigma}^2$.
            
            \State Estimate $\mu_{\mathrm{lower}}$ via EBI with confidence $1-\eta_d$.
            
            \If{$\mu_{\mathrm{lower}} \geq \mu_d$}
                \State Additional $\{x_j\}^{p-1}_{j=0}$ constructing and independent EBI verification.
                \If{$\mu_{\mathrm{lower}} \geq \mu_d$}
                    \State \Return Successful.
                \Else
                    \State \Return Failure.
                \EndIf
            \Else
                \State $p \gets np$.
            \EndIf
            
            \State $i \gets i+1.$
        \EndWhile
        
        \State The DNN needs to be retrained.
    \end{algorithmic}
\end{algorithm}

In general, the proposed approximate NMPC framework can be summarized as a trial-and-error approach, as shown in Algorithm \ref{algorithm2}.
\begin{algorithm}[h]
    \caption{Framework of Approximate NMPC}
    \label{algorithm2}
    \begin{algorithmic}[1]
        \Require Original NMPC, $\overline{e}_{\mathrm{pre}}$, $\eta_d$, $\mu_d$, and iteration limit $j_{\max}$.
        
        \State $j \gets 0$.
        
        \State Tighten the NMPC constraints using $\overline{e}_{\mathrm{pre}}$.
        
        \While{$j < j_{\max}$}
            \State Construct $\Xi$ using the tightened NMPC.
            
            \State Train the DNN to approximate the NMPC.
            
            \State Estimate $\mu_{\mathrm{lower}}$ using Algorithm~\ref{algorithm1}.
            
            \If{$\mu_{\mathrm{lower}} < \mu_d$}
                \State Enlarge $\Xi$ or increase the DNN size.
            \Else
                \State \Return Successful.
            \EndIf
            
            \State $j \gets j+1$.
        \EndWhile
        
        \State Relax $\mu_d$, $\eta_d$, or $\overline{e}_{\mathrm{pre}}$.
    \end{algorithmic}
\end{algorithm}
\FloatBarrier 
\section{Numerical Example}
\label{section:examp}
This section validates the proposed approximate NMPC framework through a nonlinear constrained control benchmark example. The NMPC is implemented in Python with CasADi \cite{casadi} and IPOPT \cite{WACHTER2006}, and the approximate NMPC is implemented in Python with PyTorch. We construct and implement the NMPC and approximate NMPC in a Windows 10 computer with an Intel Core i5-14600K processor and a NVIDIA GeForce RTX 4070 GPU.

\subsection{System and Original NMPC}
Consider a discrete nonlinear system,
\begin{equation}
\begin{aligned}
&x^+_{1}=x_1+T(x_{2}+u(\mu+(1-\mu) x_{1})) ,\\
&x^+_{2}=x_2+T(x_{1}+u(\mu-4(1-\mu) x_{2})) ,
\end{aligned}
\end{equation}
where $\mu=0.8$ and $T=0.1$.

This system is a discrete version of the example in \cite{CHEN19981205} which is unstable and uncontrollable but stabilizable.

The control and state constraints are as follows:
\begin{align*}
&\mathcal{U}=\left\{u \in \mathbb{R}^{1} \mid-2.0 \leq u \leq 2.0\right\},\\
&\mathcal{X}=\{x\in\mathbb{R}^2|-2.0\leq x_1,x_2\leq2.0\}.
\end{align*}
\indent The weighting matrices of the cost function are selected as
\[Q=\mathrm{diag}(0.5,0.5),\quad R=1.\]
\indent Following Lemma \ref{terminalcontrollaw}, we obtain the terminal state feedback control gain $K = [-1.4919  \  -4.4429]$, the terminal penalty matrix $P = \begin{bmatrix}
16.265091&13.352488\\13.352488&46.230942\end{bmatrix}$ and the terminal constraint $\mathcal{X}_f  \ \mathrm= {\  \{\mathit{x}\in \mathbb{R} ^\mathrm2 | \mathit{x^\top  Px} \leq \mathrm{8} \}}$. We solve the optimization problem in a receding horizon fashion with a sample time $t_\delta = 0.1s$ and $N_p = 5$.

\subsection{NMPC with Tightened Constraints}
\label{NMPCtightenedconstraintexample}
For the controlled system $f$, the Lipschitz constants $v_1,v_2$ are estimated using the spectral norm of $\frac{\partial f}{\partial x}$ and the $2$-norm of $\frac{\partial f}{\partial u}$ within $\mathcal{X}\times\mathcal{U}$, yielding $v_1=1.20143,v_2=0.26833$. To satisfy Theorem \ref{AsymptoticStability}, let $\overline{e}_{\mathrm{pre}}=2.60\times10^{-4}$. 

By Theorem \ref{tightenU} and Theorem \ref{tightenX}, since $\mathcal{U}$ and $\mathcal{X}$ are convex, linear scaling is applied to obtain the tightened constraints $\mathcal{U}'$ and $\mathcal{X}_i'$, with $\beta_1=0.99986$ and $\beta_{i|2}=1-1.74\times10^{-4}(1.20143^i-1),\ i\in\mathbb{I}_{0:N_p-1}$. Furthermore, $\beta_3=0.997$ by Theorem \ref{tightenX_f}.

\begin{figure}[!tbp]
\centering

\centering    
\includegraphics[width=0.7\linewidth]{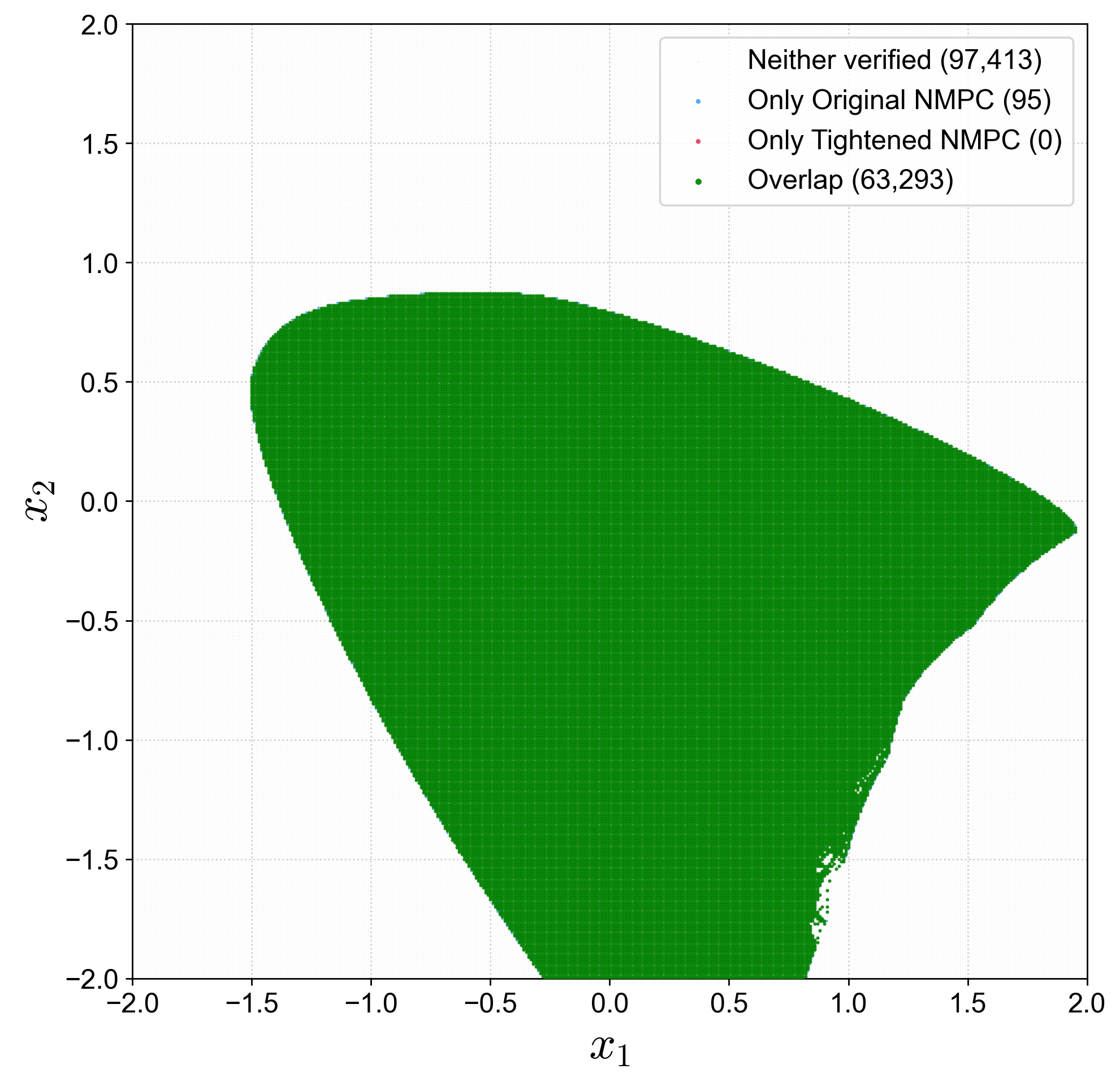}
\label{fig:subfig13}
\caption{Approximation of the feasible region via grid sampling. Only $95$ points become infeasible due to constraint tightening. Therefore, the feasible region is almost entirely covered by the green points.}
\label{fig:FeasibleRegion}
\end{figure}

Constraint tightening inevitably leads to a tightening of the feasible region of the problem. To evaluate the impact, we uniformly sample initial states from $\mathcal{X}$ at intervals of $0.01$ along each dimension and compute the solutions to Problem \ref{prob1} and Problem \ref{prob2}, respectively. The results are shown in Figure \ref{fig:FeasibleRegion}.

From Figure \ref{fig:FeasibleRegion}, nearly all feasible solution points of the two NMPCs coincide on the grid with a spacing of $0.01$, indicating that the constraint tightening does not significantly affect the feasible region of the original problem. 

\subsection{Training the Approximate NMPC}
This subsection presents training details of the DNN.

The dataset $\Xi$ comprises $6,000,000$ state-control pairs $(x,\boldsymbol{u}^*)$ generated by solving Problem~\ref{prob2} for states sampled from $\mathcal{X}$. Dataset $\Xi$ is divided into training and test sets with a ratio of $97:3$.

The DNN is designed with $8$ hidden layers of \([512,\allowbreak 512,\allowbreak 512,\allowbreak
512,\allowbreak 256,\allowbreak 256,\allowbreak 256,\allowbreak 256]\) neurons, using sinusoidal activation \cite{SIREN} as the activation function and HardTanh for the output layer. We choose the weights $\lambda_1,\lambda_2,\lambda_3$ of the loss function described in Section \ref{DNNdesign} as $1,1,0.01$ and the scale parameter $\omega$ as $2.6\times 10^{-4}$. The AdamW \cite{adamw} optimizer is used with an initial learning rate of $1\times10^{-4}$.
\begin{remark}
    There exists a trade-off between the prescribed approximation error and the network size. A smaller error bound requires higher approximation accuracy and thus a larger network, increasing computational complexity. However, for GPU-based edge platforms, the network size in this experiment remains computationally affordable, and DNN-based controllers can benefit greatly from GPU hardware acceleration.
\end{remark}

\subsection{Approximation Error Verification}
The approximation error verification scheme in Section \ref{estimateframework} is applied to estimate the lower bound of the safety probability. Additionally, we compare our method with that of \cite{Hertneck2018} to demonstrate the advantages.



The desired safety probability and significance level are set as $\mu_d=0.90$ and $\eta_d=1\times10^{-5}$, respectively. We randomly sample $100,000$ feasible states $x_j$ from $\mathcal{X}$ 
and compute the corresponding $\mathcal{E}_{\mathrm{cl}}
(x_j)$, yielding $\hat{\mu}=0.90951$ and $\hat{\sigma}^2=0.0823$. The lower bound of actual safety probability is estimated using EBI and the method of \cite{Hertneck2018}, resulting in $\mu_\mathrm{lower}=0.90474$ and $\mu_{\mathrm{lower},H}=0.90170$, respectively.

Furthermore, when the number of samples is reduced to \(10,000\), \(\mu_{\mathrm{lower}}\) and \(\mu_{\mathrm{lower},H}\) decrease to \(0.89281\) and \(0.88505\), respectively. With only \(1,000\) samples, they further decrease to \(0.84188\) and \(0.836097\), respectively.

These results indicate that the method in \cite{Hertneck2018} requires more samples to achieve a comparable estimation accuracy and yields a more conservative lower bound, making it less suitable for rapid network verification.

\subsection{Closed-loop Performance}

This subsection presents the control performance and computation time of the approximate NMPC with dual-mode law (see Theorem \ref{AsymptoticStability}) and compares them with those of the original NMPC. 

We select the corners of the feasible region from Section \ref{NMPCtightenedconstraintexample} as the initial states, namely $(-1.34,0.74)$, $(0.85,-1.7)$, $(-0.4,-1.8)$. We adopt $\|x\|\leq1\times10^{-4}$ as the convergence criterion for the trajectory.

\begin{figure}[!tbp]
\centering
\subfigure[Closed-loop trajectories]{

\centering    
\includegraphics[width=0.5\linewidth]{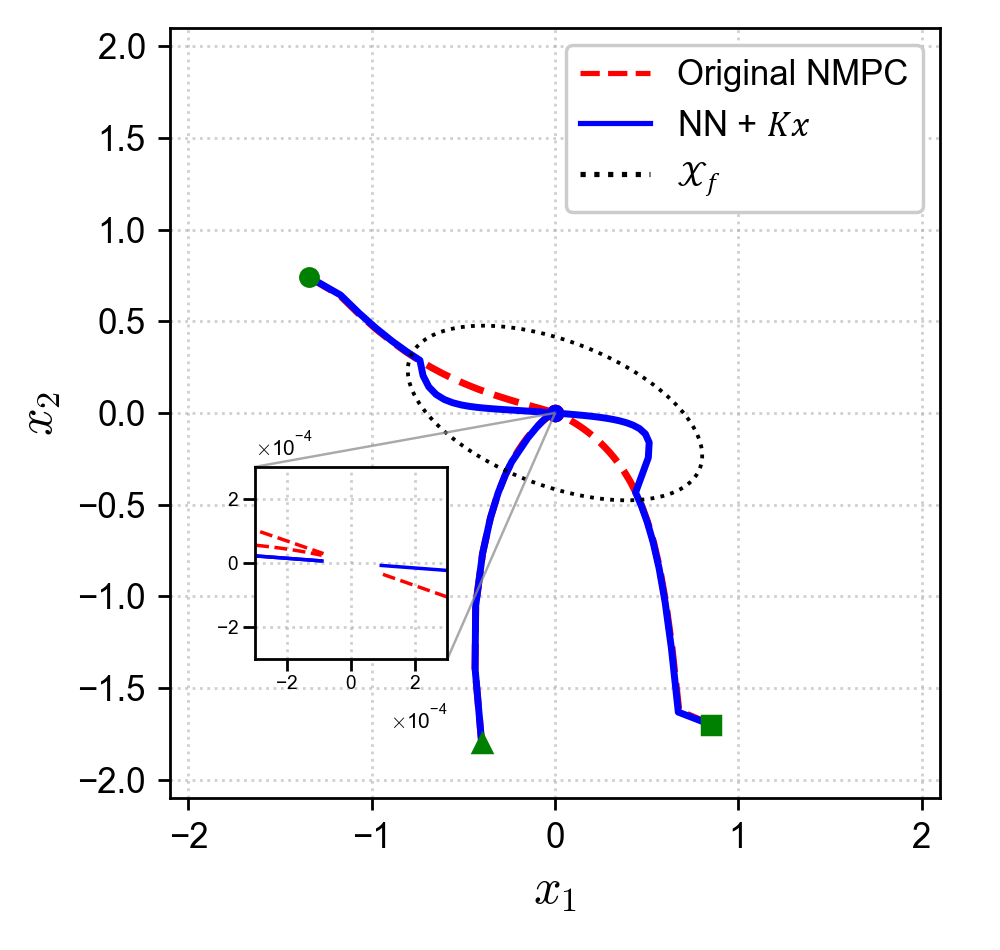}

}\subfigure[Control inputs (local)]{
\centering    
\includegraphics[width=0.5\linewidth]{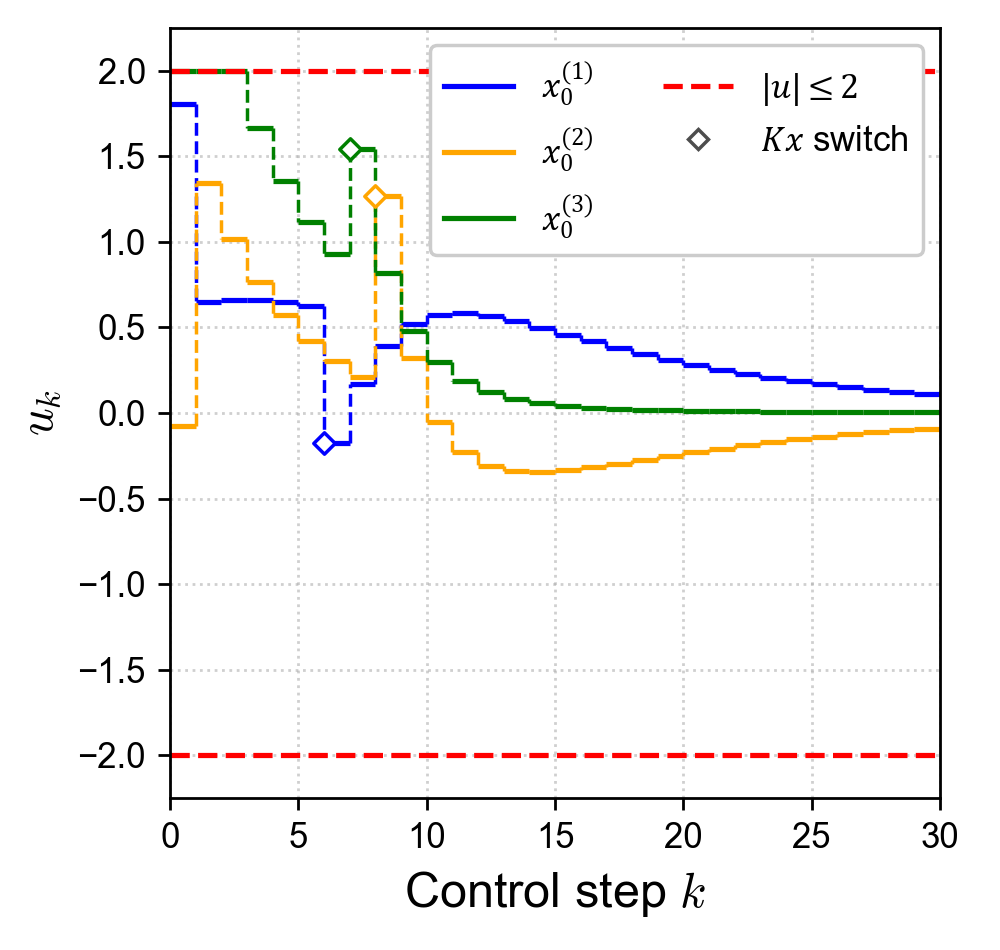}
}
\caption{Comparison of the closed-loop trajectories and control inputs. The closed-loop trajectories nearly coincide outside $\mathcal{X}_f$ but diverge after entering it due to the activation of the terminal feedback law.}
\label{fig:ANMPCnoerrorNN1}
\end{figure}

\begin{figure}[!tbp]
\centering
\includegraphics[width=0.8\linewidth]{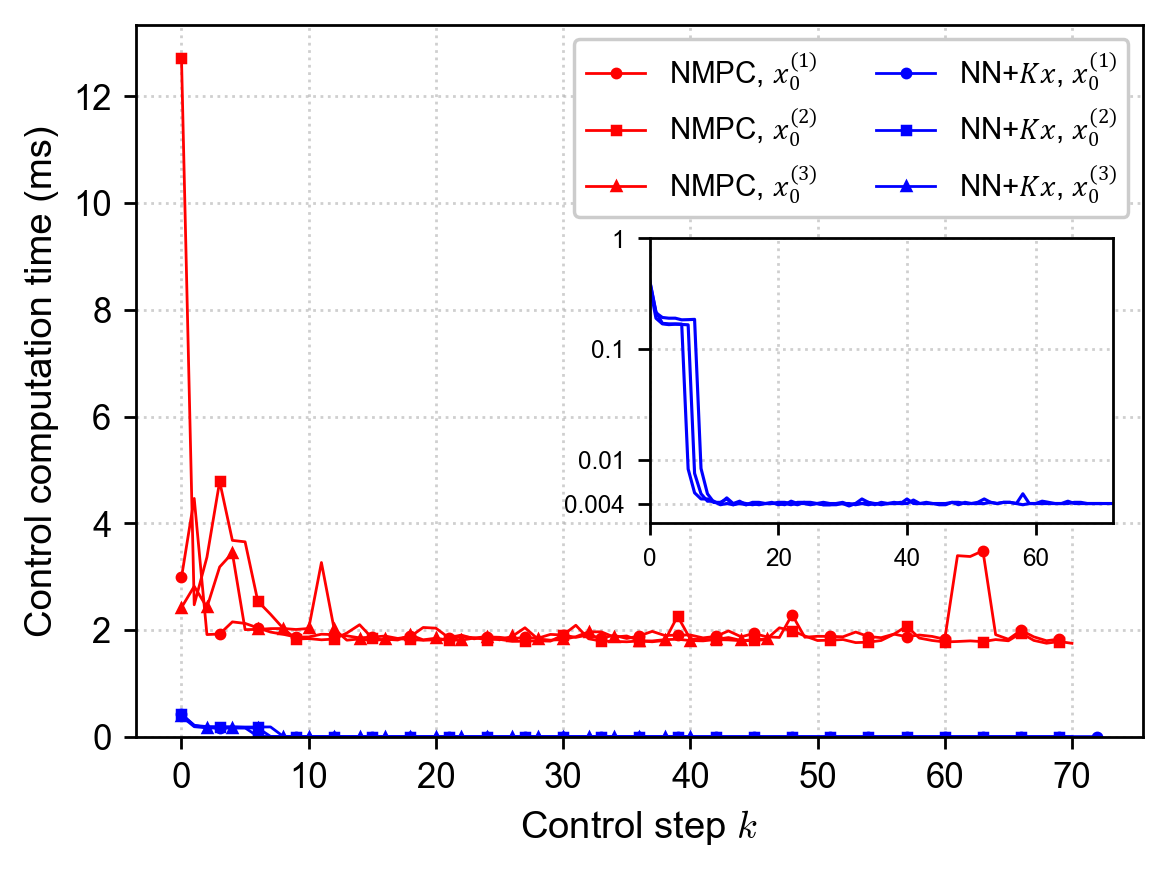}
\caption{Comparison of per-step computation time. The original NMPC has higher average computation time. }
\label{fig:ANMPCComputationTime}

\end{figure}

The closed-loop trajectories for each initial state are computed separately for the original NMPC and for the
approximate NMPC. The results are shown in Figure \ref{fig:ANMPCnoerrorNN1} and Figure \ref{fig:ANMPCComputationTime}.

In Figure \ref{fig:ANMPCnoerrorNN1}, all trajectories converge to the equilibrium, and all control inputs lie within $\mathcal{U}$. The minimum distances to the constraint boundary are $0.19522,0.6578,0.00028$, respectively.

As shown in Figure \ref{fig:ANMPCComputationTime}, the approximate NMPC significantly reduces the computation time. Across three closed-loop trajectories, the average computation times of the NMPC and approximate NMPC are $2.064$~ms and $0.0278$~ms, respectively. The original NMPC requires an additional $127.6256$~ms per trajectory on average, with a maximum computation time approximately $12.2845$~ms higher than that of the approximate NMPC.

\subsection{Safety Performance}
\begin{figure}[!tbp]
\centering
\includegraphics[width=0.8\linewidth]{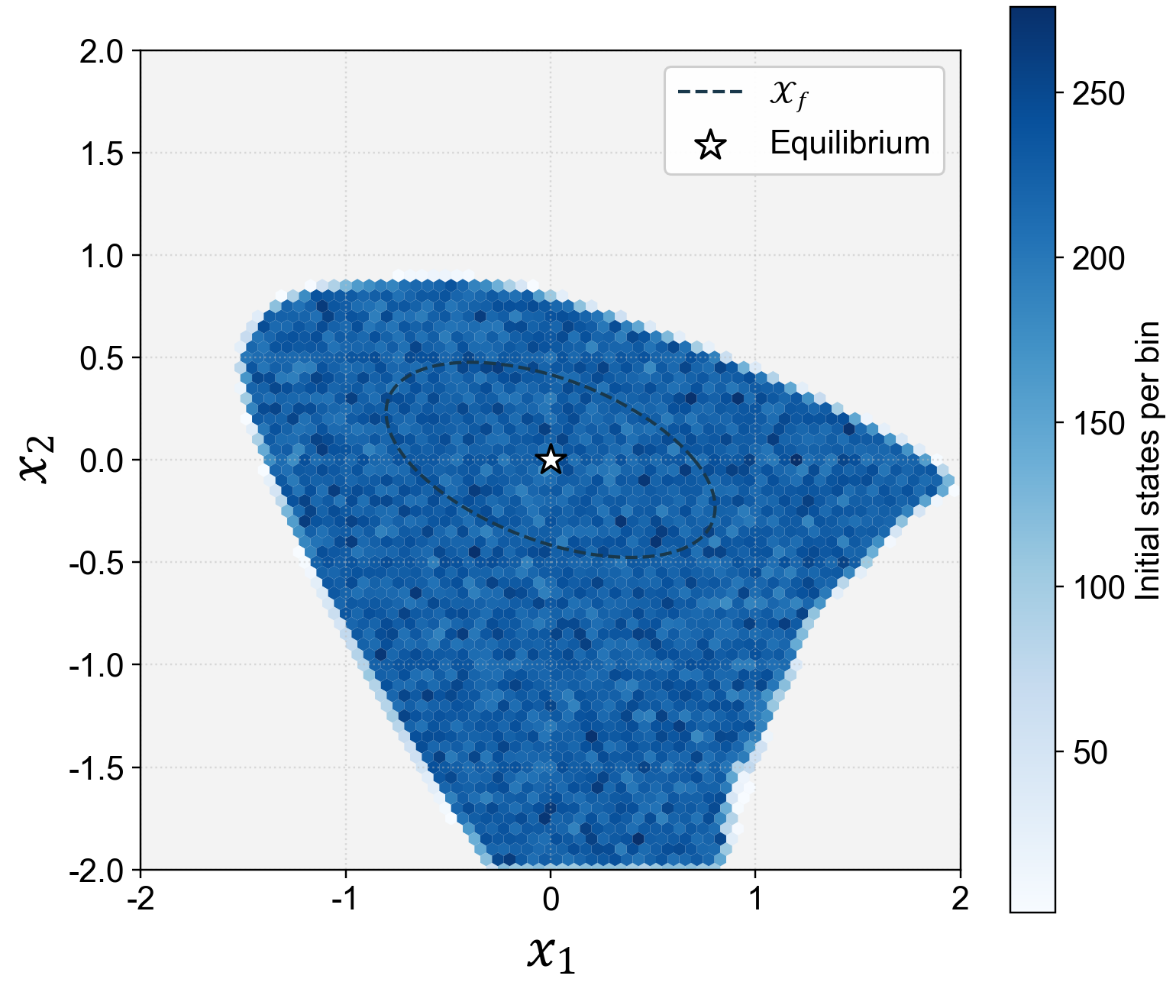}
\caption{The distribution of the $500,000$ sampled initial states.}
\label{fig:InitialStates}

\end{figure}

This subsection evaluates the actual safety probability and the empirical stability probability of the approximate NMPC. To this end, we randomly sample $500,000$ feasible initial states over $\mathcal{X}$, run the approximate NMPC until convergence to the equilibrium, and compute the safety probability from the samples. The sampled initial states distribution is shown in Figure \ref{fig:InitialStates}.

According to experimental statistics, all $500,000$ trajectories converged without constraint violations, and the observed minimum distances to the state and control constraint boundary are $1.54\times10^{-5}$ and $2.80\times10^{-4}$, respectively. These observations support the effectiveness of the approximate controller and suggest that the theoretical safety guarantee is conservative.

\section{Conclusions}
\label{section:con}
This paper proposed a learning-based approximate NMPC framework with probabilistic guarantees of constraint satisfaction and stability. Constraint tightening was employed to ensure robustness against a prescribed maximum approximation error, while an iterative network training and error verification procedure was developed to provide a probabilistic guarantee that the approximation error does not exceed the prescribed maximum throughout the entire closed-loop trajectory. Experimental results demonstrated comparable closed-loop performance to that of the original NMPC at a lower computational cost, while providing empirical evidence of closed-loop safety and stability, highlighting the framework's potential for computationally demanding constrained nonlinear control problems.

However, the stability conditions remain conservative, and more complex tasks may require tighter constraints, larger networks, and additional training data. Moreover, control law switching may introduce a transient shock to the system.

\section{Acknowledgments}
Partially supported by the Natural Science Foundation of China under Grant 62473167, the Natural Science Foundation of Jilin Province 20260201092GX.

\bibliographystyle{unsrt}  
\bibliography{references}  

\end{document}